\documentclass[conference]{IEEEtran}
\usepackage{tabularx}
\usepackage{listliketab}

\usepackage{lipsum}
\usepackage{algorithm}

\usepackage{algorithmic}

\usepackage[T1]{fontenc}
\usepackage{cite}
\usepackage{url}
\usepackage{amsmath}
\usepackage[cmintegrals]{newtxmath}
\usepackage[caption=false,font=footnotesize]{subfig}
\usepackage{graphicx}
\usepackage{epstopdf}
\usepackage{xcolor}

\usepackage{enumitem}
\begin{document}
%
\title{Context-aware Cluster Based Device-to-Device Communication to Serve Machine Type Communications}

\author{\IEEEauthorblockN{Ji Lianghai, Liu Man, Hans D. Schotten}
\IEEEauthorblockA{Chair of Wireless Communication, University of Kaiserslautern, Germany\\ $\lbrace$ji,manliu,schotten$\rbrace$@eit.uni-kl.de}}

%


\IEEEoverridecommandlockouts
\IEEEpubid{\makebox[\columnwidth]{\copyright~Copyright 2017
		IEEE \hfill} \hspace{\columnsep}\makebox[\columnwidth]{ }}

\maketitle

\begin{abstract}
Billions of Machine Type Communication (MTC) devices are foreseen to be deployed in next ten years and therefore potentially open a new market for next generation wireless network. However, MTC applications have different characteristics and requirements compared with the services provided by legacy cellular networks. For instance, an MTC device sporadically requires to transmit a small data packet containing information generated by sensors. At the same time, due to the massive deployment of MTC devices, it is inefficient to charge their batteries manually and thus a long battery life is required for MTC devices. In this sense, legacy networks designed to serve human-driven traffics in real time can not support MTC efficiently. In order to improve the availability and battery life of MTC devices, context-aware device-to-device (D2D) communication is exploited in this paper. By applying D2D communication, some MTC users can serve as relays for other MTC users who experience bad channel conditions. Moreover, signaling schemes are also designed to enable the collection of context information and support the proposed D2D communication scheme. Last but not least, a system level simulator is implemented to evaluate the system performance of the proposed technologies and a large performance gain is shown by the numerical results.
\end{abstract}


%
\IEEEpeerreviewmaketitle

\section{Introduction}
As one of the new emerging services, Machine Type Communication (MTC) \cite{lab1}\cite{lab2}  is considered by many researchers and experts as an important service in the coming fifth generation (5G) cellular  network. In the 4G network, legacy LTE-A network was designed to offer high data rate, low latency, high spectrum efficiency and high system capacity. Thus, 4G network experiences technical challenges to offer MTC services since different considerations and requirements are posed on MTC services, e.g., a massive deployment of devices, a low device complexity and a long device battery life. \\
In the third Generation Partnership Project (3GPP), related studies are conducted to evolve legacy network to meet the requirements of MTC services. For instance, a new user equipment (UE) category (UE category ``0'') is introduced to reduce device complexity and power consumption in \cite{lab2}. In addition, an extended discontinuous reception (DRX) is considered in \cite{lab4} to reduce battery consumption where longer sleep cycles are exploited and optimized for delay-tolerant MTC applications. Moreover, removal of the power amplifier is proposed in \cite{lab2} to reduce device cost. However, uplink coverage is reduced in this case due to a lower maximal transmission power. Another challenge faced by MTC is the extra penetration loss of 20 dB due to the deep-in-door deployment of users (UEs) \cite{lab5}. These MTC UEs located deep-in-door are referred as remote UEs due to their bad cellular channel conditions. Compared with base station (BS), since a lower transmission power is available at UE, it is more challenging to maintain the network coverage in uplink. Existing solution to maintain uplink coverage is to either use narrow band transmission or exploit massive transmission time interval (TTI) bundling \cite{lab5}. Though both schemes help in enhancing the MTC availability, a large resource usage at system level and a battery drain at device level are deduced. In another work \cite{lab6}, it is proposed to exploit the 3GPP defined relay nodes (RNs) to improve coverage for MTC. However, RNs are usually exploited for mobile broadband service and they normally locate in areas with busy human activities. Thus, deployment of RNs in deep-in-door and rural area cannot always be assumed.\\
As one of the critical technical enablers for 5G cellular network, device-to-device (D2D) communication \cite{after6_1}\cite{after6_2} opens the opportunity to improve the performance of cellular networks. The motivation of exploiting D2D communication was to either offload cellular traffic to local information exchange procedure or to enable a direct D2D communication to achieve low latency \cite{lab7}. In research work \cite{lab8}\cite{lab9}, the applied scenario of D2D communication is extended to MTC services. In these work, a normal cellphone with D2D discovery and communication capability acts as a relay for other sensors. Together with academic community, 3GPP also considers exploiting cellphones as relays for MTC applications \cite{lab10}. In this approach, several drawbacks exist, as listed in the following items.
\begin{enumerate}
\item D2D discovery procedure is carried out every time when a remote UE is paged or has uplink date in its buffer to transmit. In this way, extra power consumption is deduced.
\item Since the D2D pairing is performed in a distributed manner without any help from BS, it brings a lost in global awareness. For instance, once a device is paired with one relay device, it can not get served by other potential relay devices though they might be better choice. 
\item Last but not least. presence of cellphones in a deep-in-door scenario and rural area can not always be expected.
\end{enumerate}
Instead of using cellphones as relays, one of the MTC devices is appointed as an aggregator in \cite{lab11}, where full knowledge of Channel Quality Indicators (CQIs) is necessary to be available in BS to set up D2D pairs properly. Thus, if a cluster consists of $M$ devices, then BS needs to collect $\frac{M\times (M-1)}{2}$ CQIs information with each CQI representing the channel situation between two devices. This requirement leads to a very cumbersome signaling procedure with high power consumption and signaling overload.\\
In this work, we inspect on how to improve the availability and battery life of MTC UEs by exploiting D2D communication. MTC UEs are assumed to be statically distributed inside buildings. In the proposed approach, MTC UEs located deep-in-door have the opportunity to set up D2D links with relay MTC UEs. Compared with the existing work in literature, our proposal contributes to the following aspects.
\begin{itemize}
\item Signaling schemes with low overhead are provided to support the exploitation of D2D communication in the considered scenario.
\item The signaling schemes enable an efficient collection of context information.
\item With the help from the collected context information, BS can optimize the transmission mode (TM) of each user and improve the system performance.
\end{itemize}
Our work is organized as follows. At beginning, a cluster based transmission mode selection (TMS) scheme is proposed in Sect.~\ref{tsm}. After that, corresponding signaling schemes are provided in Sect.~\ref{rle} to support the proposed D2D communication with a low signaling load. In Sect.~\ref{em}, methodology used to evaluate the system performance of MTC services is stated with details. Simulation results are given in Sect.~\ref{nr} where a large gain in terms of availability and battery life can be seen. Finally, we conclude this work in Sect.~\ref{conc}.
\begin{figure}[!b]
\centering
\includegraphics[width=0.5\textwidth]{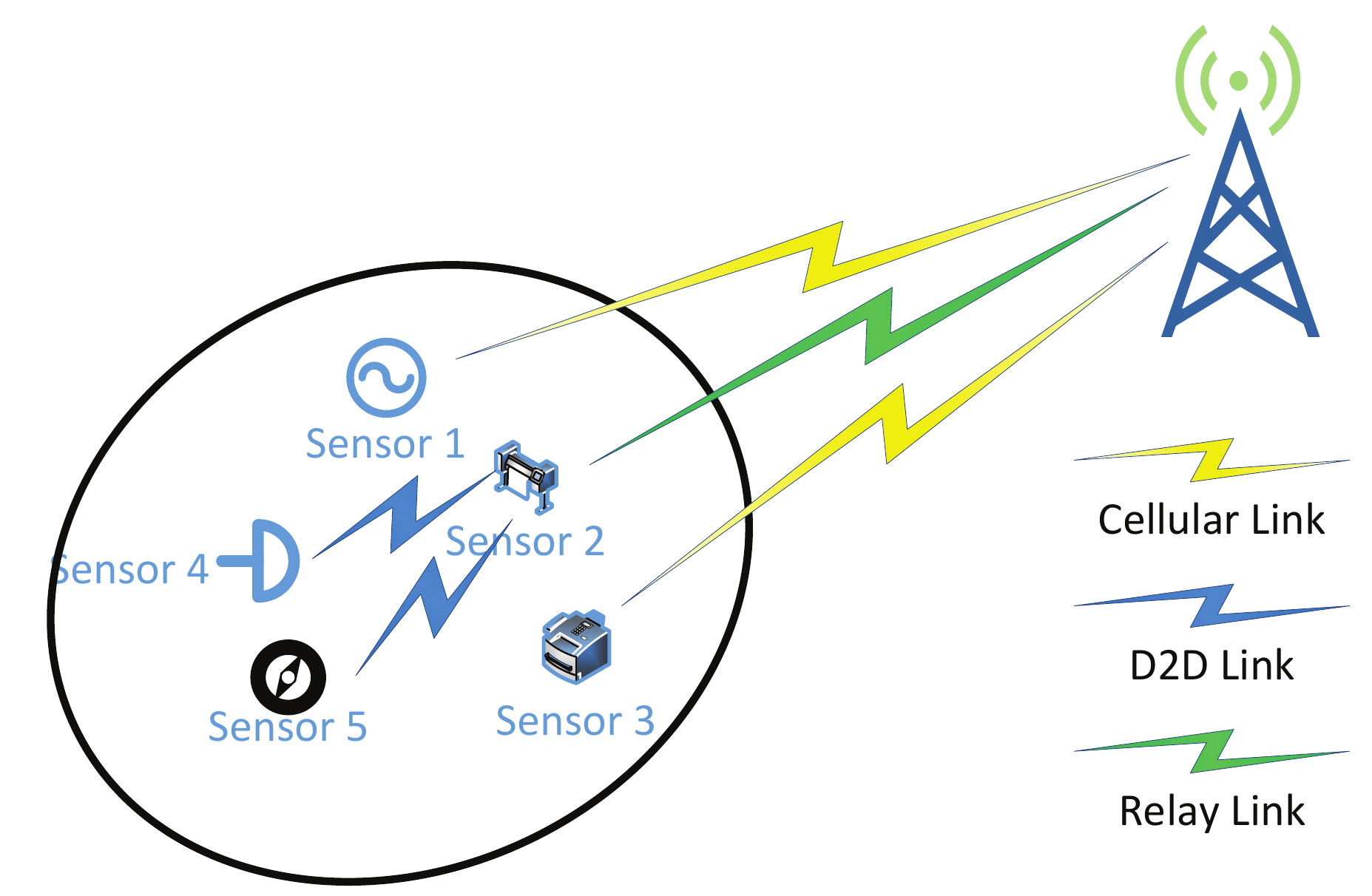}
\caption{scenario description}
\label{sd}
\end{figure}
%
%
\section{Transmission mode selection based on virtual sectors}\label{tsm}
As mentioned before, D2D communication is exploited in this work to facilitate the uplink reports from remote UEs. In this scheme, three different transmission modes exist, as following:
\begin{itemize}
\item cellular transmission mode, in which the devices upload their reports to BS with cellular links;
\item relay transmission mode, in which the devices are configured by network to relay the reports from remote UEs and meanwhile transmit their own reports to BS;
\item D2D transmission mode, in which the remote UEs transmit their reports to relay UEs.
\end{itemize}
In order to adapt to any system changes in real time, MTC UEs are dynamically configured with their transmission modes by BS. Fig.~\ref{sd} provides a graphical description of this scenario. As it can be seen, sensor $\sharp4$ and sensor $\sharp5$ experience bad channel conditions for their cellular links and thus are referred as remote UEs. Meanwhile, sensor $\sharp2$ is seen by BS as an optimal relay node for sensor $\sharp4$ and $\sharp5$. Thus, D2D connections are established between sensor $\sharp2$ and sensor $\sharp4$, also between sensor $\sharp2$ and sensor $\sharp5$. After that, uplink data of sensors $\sharp4$ and $\sharp5$ are transmitted to BS through sensor $\sharp2$. Besides being a relay node for remote UEs, sensor $\sharp2$ also transmits its own packet to BS. Moreover, sensors $\sharp1$ and $\sharp3$ are configured as normal cellular UEs and they are only responsible for transmissions of their own packets.\\
From efficiency point of view, D2D communication should be applied in cases where transmitter and receiver are nearby each other. Thus, a clustering approach is required at BS to make sure that a relay UE only serves remote UEs in its proximity. Afterwards, BS needs to select proper transmission mode for each UE, taking into account of the context information (e.g., channel state information (CSI) between BS and the UE, location and battery level information). Therefore, the proposed context-aware D2D communication can be divided into two steps:
\begin{enumerate}
\item clustering devices into different groups;
\item selection of transmission mode for each UE.
\end{enumerate}
\subsection{Virtual clustering}
In this section, four different methods with different considerations are introduced to group devices. Moreover, for devices locating near the BS, their propagation losses are relatively low and thus D2D communication is not applied for these UEs. The area without application of D2D communication can be represented by a cycle with a radius of $R_{in}$.\\ 
\subsubsection{Geometrical clustering}  
\begin{figure}[!t]
\centering
\includegraphics[width=0.5\textwidth]{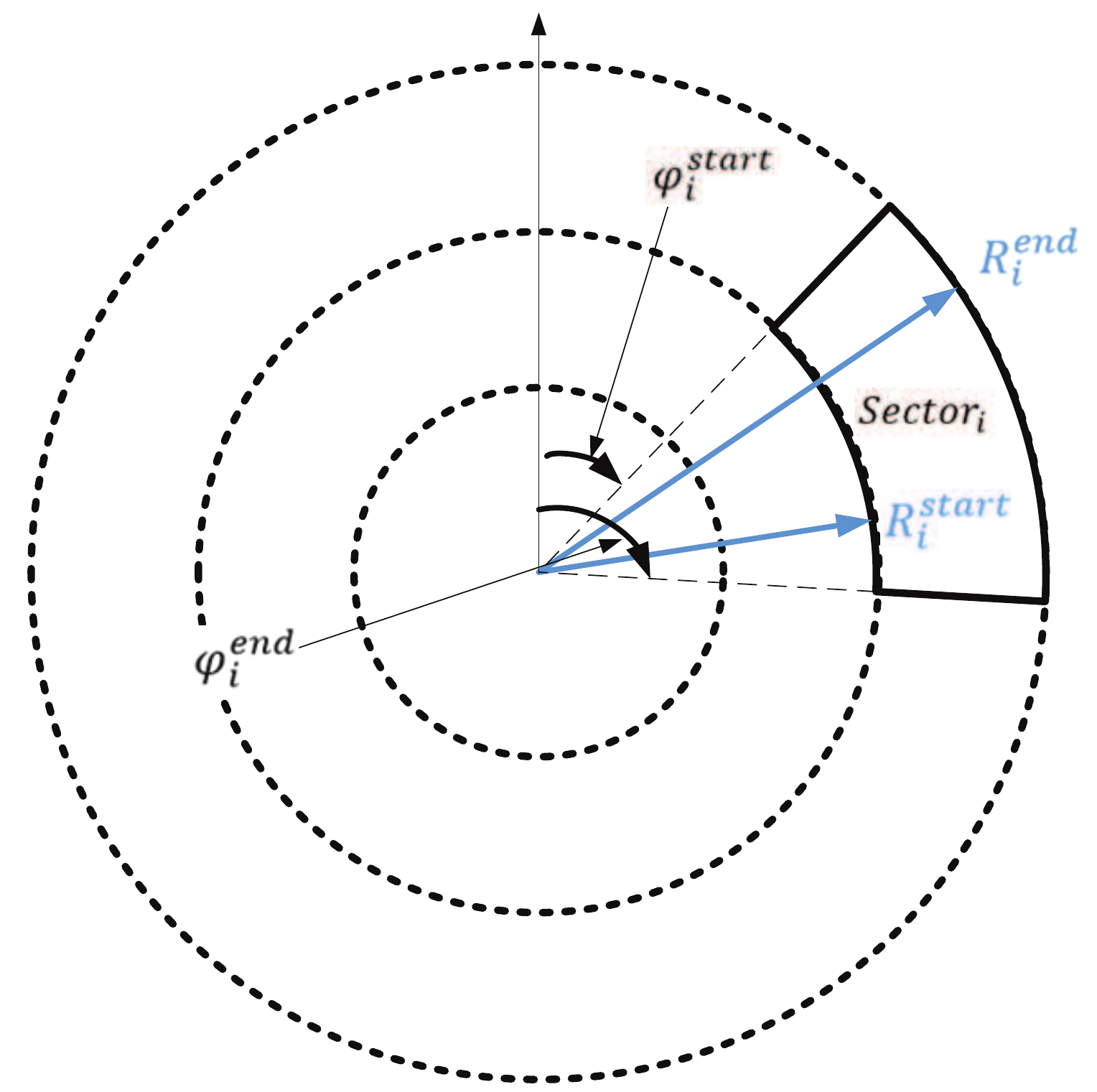}
\caption{geometrical clustering}
\label{gc}
\end{figure}
In this method, the coverage area of one BS is sectorized geometrically, as shown in Fig.~\ref{gc}. The area covered by the $i$-th cluster can be represented by a radius of $R_i$ and an angle of $\varphi_i$, as
\begin{equation}
R_i^{start}<R_i\leq R_i^{end},
\end{equation}
\begin{equation}
\varphi_i^{start}<\varphi_i\leq \varphi_i^{end}.
\end{equation}
where $R_i^{start}$ and $R_i^{end}$ represent the distances from the origin (i.e. the BS) to inner and outer circles of the $i$-th cluster. Moreover, $\varphi_i^{start}$ and $\varphi_i^{end}$ represent the reference angles between which the $i$-th cluster covers. The number of clusters covered by one BS is function of $A_{sector}$, which is the area of one cluster.
\subsubsection{K-means clustering}
K-means is one well-known clustering algorithm and its basic steps are listed below.
\begin{enumerate}
\item [(1)] Initially, we select K devices which are placed as far away as possible from each other and these K devices are considered as centroids of the K clusters.
\item [(2)] Then, another device is selected and associated to the cluster, the centroid of which has the shortest distance from the selected device.\label{stepK2}
\item [(3)]Calculate the mean coordinate of the new formed cluster and select the nearest device to the mean coordinate as the new centroid.\label{stepK3}
\item [(4)] Repeat step (2) and (3) until all devices are associated to a cluster.
\end{enumerate}
\subsubsection{Distance based clustering}
This scheme is similar to the K-means clustering algorithm, with following steps. 
\begin{enumerate}
\item [(1)] K devices are randomly selected as centroids of K clusters.\label{stepD1}
\item [(2)] Take one another device and associate it to the cluster which has the shortest distance from its centroid to this device.\label{stepD2}
\item [(3)]Repeat step (2) until all devices are associated to a cluster.
\end{enumerate}
It can be noticed that, the centroids are selected randomly from the data set and these centroids are not updated during the operation of this algorithm. These are the two difference of this scheme compared with the K-means clustering method.
\subsubsection{Distance plus CSI based clustering}
In this scheme, not only the location information of UEs are considered, but also the CSI between each UE and the BS. The difference compared with distance based clustering method is that, the K centroids are selected from devices which have cellular SNR values higher than a pre-defined threshold.
\subsection{Transmission mode selection}
The task of TMS is to configure devices in each cluster so that each device is aware of the transmission mode it should apply for the uplink report. Multiple context information are taken into account in this work to achieve an efficient TMS algorithm. As mentioned before, the battery life requirement of MTC devices can be up to 10 years \cite{lab5}. For UEs who cannot meet the battery life requirement, D2D communication is exploited. The equation below describes the condition of remote devices whose battery life requirement cannot be met by cellular links:
\begin{equation}\label{noFul}
\frac{BC_{(i,j)}}{EC_{(i,j)}}<BL_{threshold}.
\end{equation}
The requirement of battery life is denoted by $BL_{threshold}$. $BC_{(i,j)}$ refers to the battery capacity of user-$j$ in cluster-$i$ and $EC_{(i,j)}$ is the energy consumption by using cellular link for a time unit of $\bigtriangleup t$. $\bigtriangleup t$ is the time difference between any two successively TMS update commands for user-$j$. With a smaller value of $\bigtriangleup t$, the transmission modes of UEs are updated with a higher frequency and network can respond to condition changes in a more timely manner. On the other hand, a smaller value of $\bigtriangleup t$ also introduces a higher signaling load. Thus, in order to achieve a compromise in between the efficiency and the signaling load, $\bigtriangleup t$ can have a value ranging from several hours to several days, depending on traffic models of UEs. Moreover, users which cannot reach BS with cellular links can be assumed to have an infinite value of energy consumption for $\bigtriangleup t$. Thus, these users also fulfill the inequality in Eq.~(\ref{noFul}) and D2D communication is also applied to improve their availability.\\
If some UEs are classified as remote UEs in a cluster, BS checks whether some UEs in the same cluster fulfill the following conditions for being relays: 
\begin{equation}\label{Ful}
\frac{BC_{(i,j)}}{EC_{(i,j)}}> BL_{threshold},
\end{equation}
\begin{equation}\label{PL}
SNR_{(i,j)}^{cellular}\geq SNR_{threshold}.
\end{equation}
Eq.~(\ref{Ful}) represents the condition that user-$j$ in cluster-$i$ can meet the battery life requirement by using cellular transmission. In other words, this user has enough battery capacity to serve as a relay for other remote UEs in cluster-$i$. In Eq.~(\ref{PL}), $SNR_{(i,j)}^{cellular}$ is the SNR value of the cellular link and $SNR_{threshold}$ is a threshold value to check whether the channel condition of the cellular link is good enough. For example, a value of 3 dB is used in Sect.~\ref{nr} as $SNR_{threshold}$. With a higher value for this parameter, UEs with better cellular link CSI are considered as relay UEs and thus a higher spectral efficiency can be achieved for the relay link. However, a higher value of $SNR_{threshold}$ means less feasible relay UEs and there is a higher risk that no relay UE exists in a cluster.\\
Once the BS obtains the list of feasible relay UEs in one cluster, BS picks up one relay UE and sends the D2D setup command to both the relay UE and remote UE(s). Upon receiving the D2D setup command, channel conditions between the relay and remote UEs are estimated to inspect if the D2D communication can contribute to a better service availability and energy efficiency. If a D2D setup procedure is successful, the established D2D link is exploited for uplink transmission of packets from the remote UE. The corresponding signaling schemes are detailed in Sect.~\ref{rle}.
\section{Radio link enablers}\label{rle}
In this section, we introduce the signaling schemes to support the proposed context-aware D2D communication. 
\subsection{TMS and D2D cluster formation}\label{signaling1}
In case either a UE is initially attached to the network or BS conducts the TMS update procedure, the signaling procedure shown in Fig.~\ref{update} will be performed to configure the corresponding UEs. Details of this signaling scheme are given below, organized by the steps shown in the figure.
\begin{figure}[!t]
\centering
\includegraphics[width=0.51\textwidth]{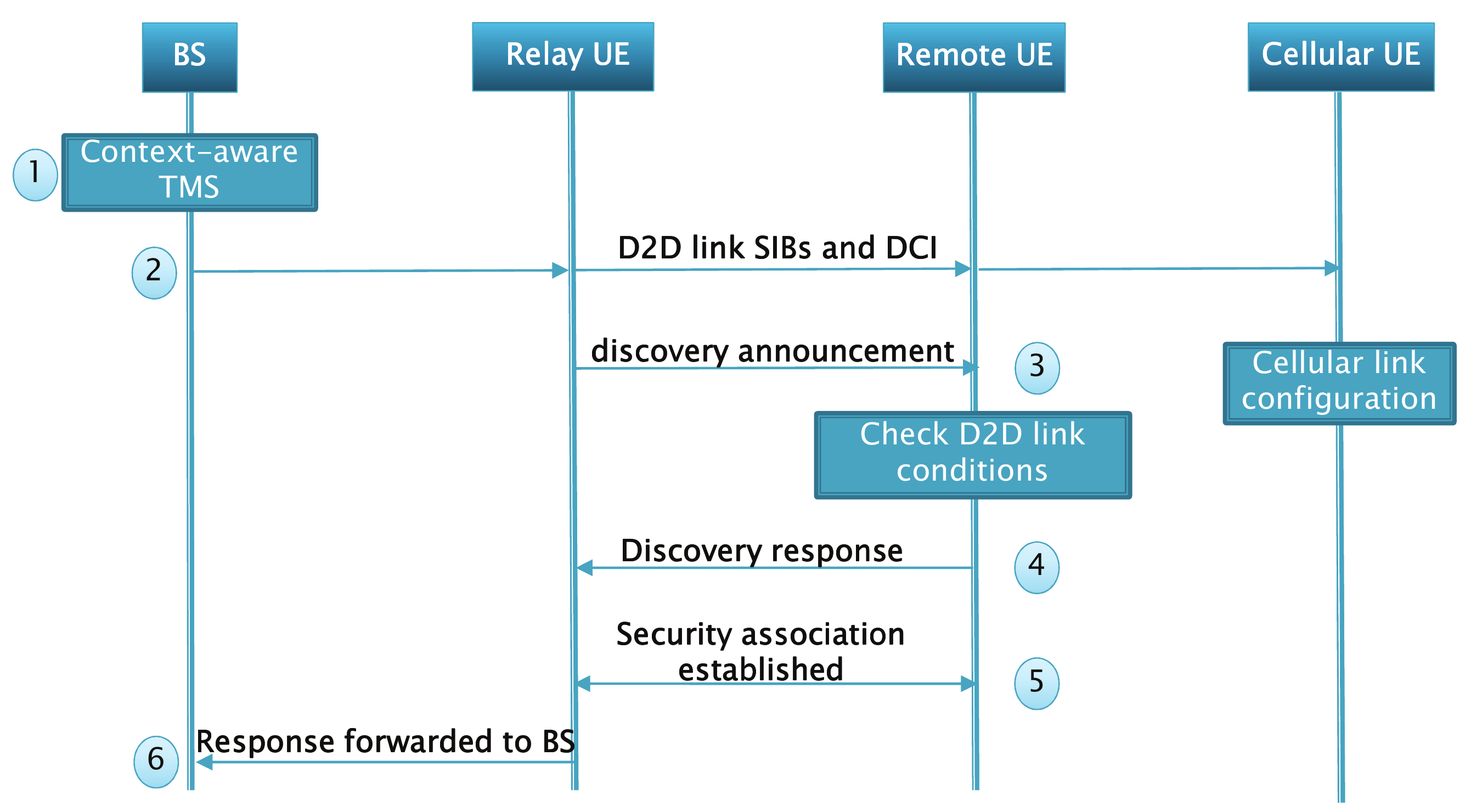}
\caption{Signaling scheme for D2D cluster formation and TMS update}
\label{update}
\end{figure}
\begin{enumerate}
\item BS collects the context information, e.g., location, battery level, traffic type and received signal reference power (RSRP) of its served MTC UEs. Besides, the context-aware clustering algorithm and TMS procedure are also performed.\label{A2start}
\item BS configures UEs with D2D link system information blocks (SIBs) for direct D2D discovery, e.g. resource used for D2D discovery. Moreover, UEs will also be informed by downlink control information (DCI) for the user specific control information, e.g. ID of the cluster to which the UE belongs, transmission mode of the UE, conditions which D2D link should fulfill and so on. In case if a UE is configured as either a relay or a remote UE, information of the other end of this D2D link are also provided.
\item For relay UE, it send D2D discovery announcement to the target remote UEs, with its own ID and IDs of the target UEs being conveyed in the announcement message. In this message, reference signals for D2D link channel estimation are also carried.\label{SD}
\item Upon receiving the D2D discovery announcement, the remote UE determines whether the request is accepted or not, based on the estimated D2D link RSRPs. And a response message of acknowledgement/non-acknowledgement is sent back to the relay UE.\label{FB}
\item If the relay UE receives an D2D acknowledgement message, a security association could be established by exchanging messages between the relay and remote UEs with security algorithms.\label{A2end}
\item The response of the D2D cluster formation is further transmitted from the relay UE to the serving BS. In case the D2D request is not accepted, the serving BS should be informed and thus it can avoid to pair the previous selected relay UE with this remote UE in future.
\end{enumerate}
If a UE is configured by BS to have cellular transmission mode, it is not involved in the D2D discovery procedure. It is also to be noticed that, when the UE is initially deployed by people, the location information, traffic type and cellular link RSRP of one MTC UE mentioned in step~\ref{A2start}) can be collected. Since MTC UEs are assumed to be static in this work, these information can be foreseen as unchanged. Besides, the actual battery level of one UE can be transmitted to BS together with the data transmission in uplink. Last but not least, if a D2D link is successfully established in step.~\ref{A2end}), configuration information of this D2D link should be stored at both the relay and remote UEs.
\subsection{D2D communication}
Once a D2D link is established according to Sect.~\ref{signaling1}, the D2D link is used for uplink packet transmission of the remote UE. The corresponding signaling scheme is shown in Fig.~\ref{communication} with details given below.
\begin{enumerate}
\item Relay and remote UEs involved in D2D communication are configured by the SIBs and DCI. These information are configured and stored at both relay and remote UEs when the D2D discovery procedure is accepted.
\item In mobile terminated case (e.g., UE is paged by network for its report), one remote UE or a group of remote UEs will be paged by BS for data report, or
\item In mobile originated case (e.g., data is available in the buffer of the UE and waits for being transmitted), a remote UE generates a data packet and tries to transmit this packet to the relay UE. This step also includes the random access procedure, D2D link connection setup procedure between the relay UE and remote UE, also D2D retransmission if an error occurs in transmission. In this step, both D2D ends should be aware of the time and frequency resource for D2D transmission.
\item After successful receiving packets from remote UE(s), relay UE replies with acknowledgment message(s) to the remote UE(s).
\item Relay UE further forwards the successfully received packets to the serving BS. This process can be performed as a normal cellular uplink transmission where a control plane (CP) connection needs to be established. Another alternative is that the received packets will be buffered in the relay UE and then transmitted together with its own packet. In this case, an advantage in power saving for relay UE can be introduced, since the relay UE only needs to wake up and perform the CP connection establishment procedure for once.
\item Upon successfully receiving packets from the relay UE, BS sends an acknowledgment message to the relay UE. 
\end{enumerate}
\begin{figure}[!t]
\centering
\includegraphics[width=0.5\textwidth]{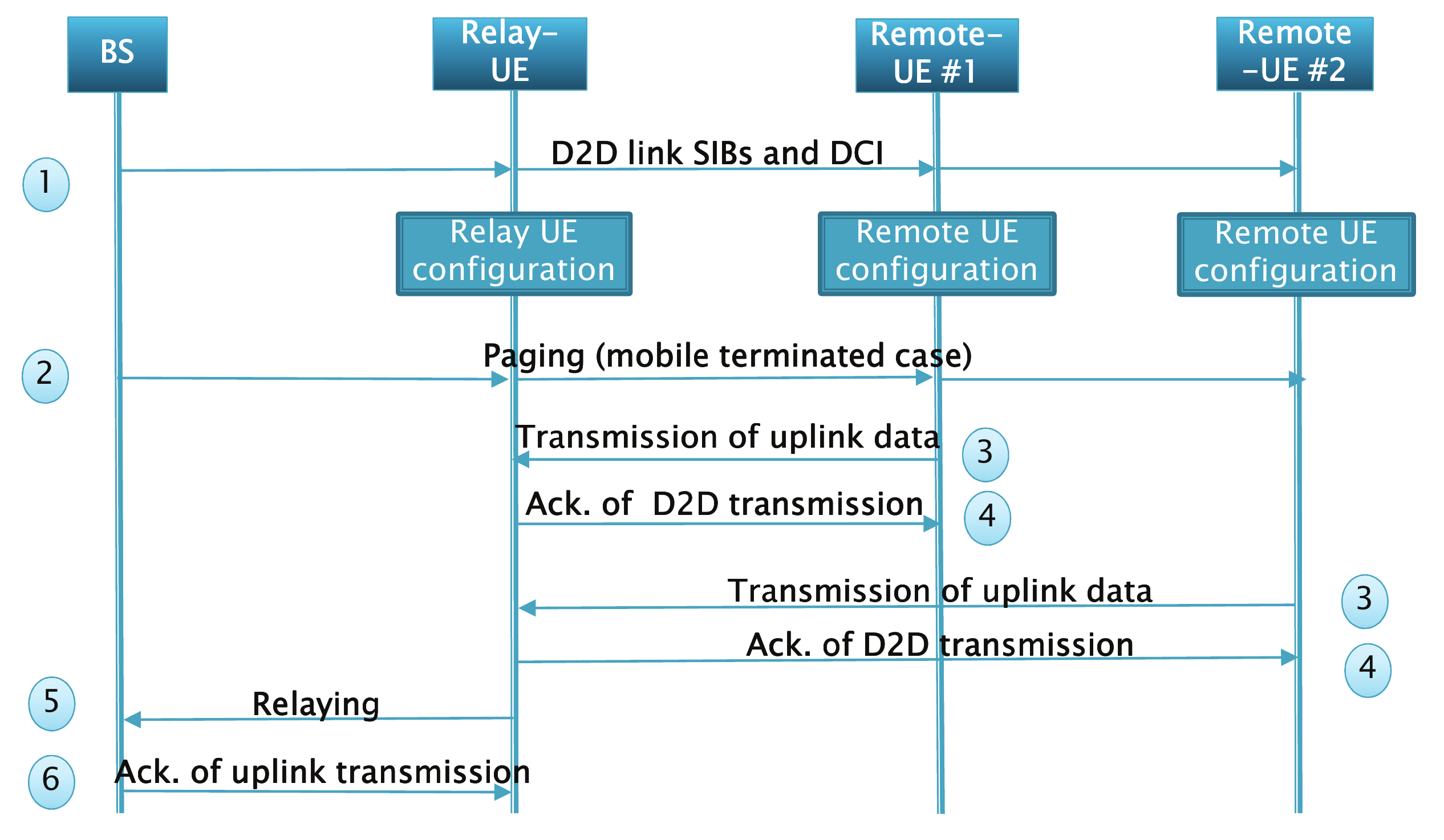}
\caption{D2D communication procedure}
\label{communication}
\end{figure}
\section{Evaluation methodology}\label{em}
In order to evaluate the proposed technology, a system level simulator is implemented in this work and aligned tightly with the real world. In this section, models used in our simulator are stated with details. Please note that, only the difference compared with ITU-R performance evaluation document \cite{ITU} are given here. For other parameters not mentioned here, they are aligned with the ITU-R document \cite{ITU}. 
\begin{figure}[!t]
\centering
\includegraphics[width=0.5\textwidth]{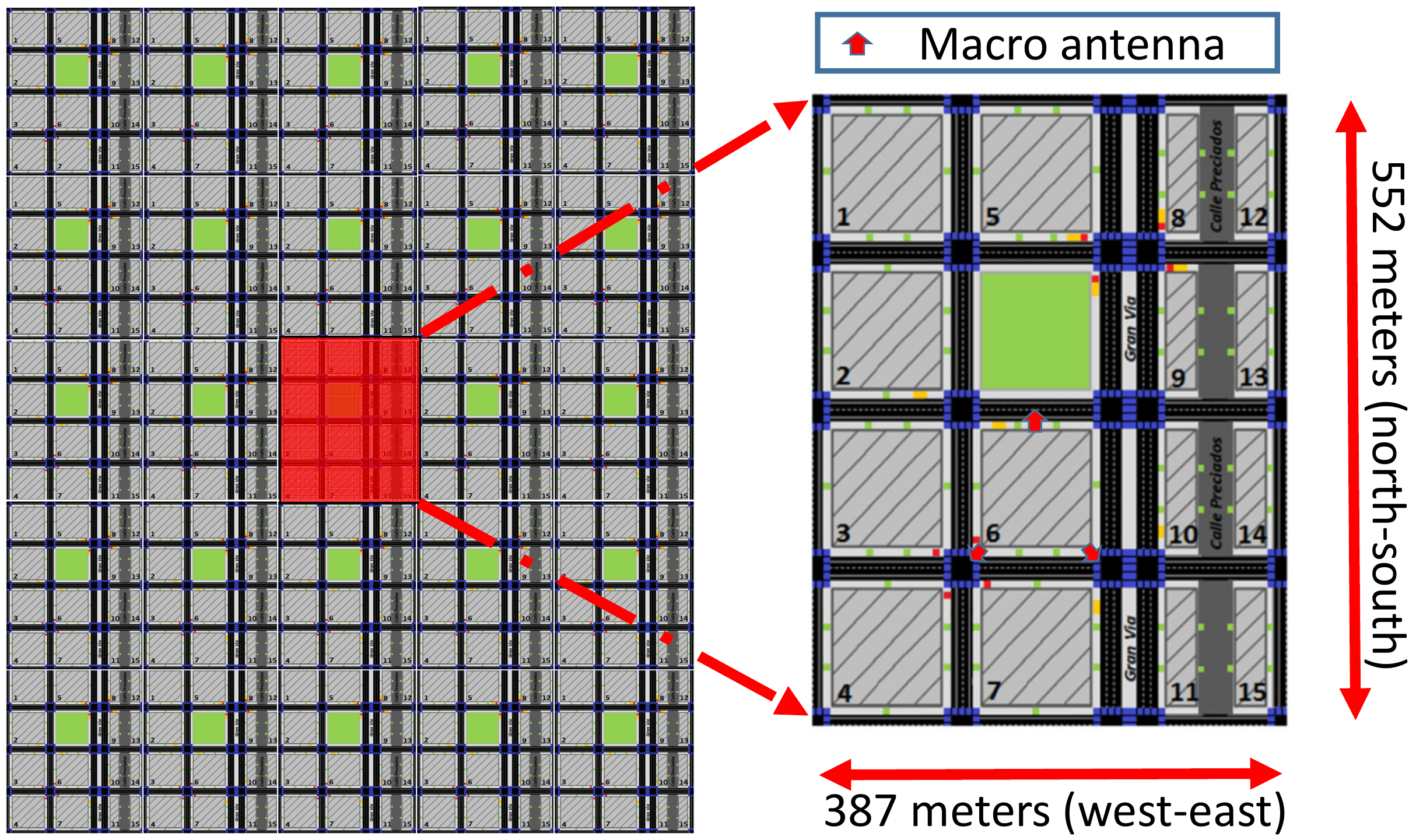}
\caption{Environment and deployment model}
\label{edm}
\end{figure}
\subsection{Environment model}
\newcommand{\tabincell}[2]{\begin{tabular}{@{}#1@{}}#2\end{tabular}}
\begin{table*}\caption{Device power consumption parameters}
\label{DC}
\begin{center}
\begin{tabular}{ |c|c|c|c| }
    \hline
     Parameter & Description & Value & Time duration if applicable \\ \hline
$P_{\textbf{tx}}$ & transmission power &  \tabincell{c}{45$\%$ PA efficiency plus 60\\mW/s for other circuitry}&  \tabincell{c}{MCS and packet\\size related}\\ \hline
    $P_{\textbf{rx}}$ & power to receive packets from remote UEs & 100 mW/s &  \tabincell{c}{MCS and packet\\ size related}\\ \hline
    $P_{\textbf{paging}}$& power to receive paging command & 100 mW/s & 10 ms
 \\ \hline
    $P_{\textbf{clock}}$& clock to obtain synchronization &100 mW/s & 10 ms \\ \hline
    $P_{\textbf{cp}}$ &  \tabincell{c}{power consumption\\during thecontrol plane\\establishment procedure}& 200 mW/s &  10 ms\\ \hline
    $P_{\textbf{sleep}}$ & power consumption in sleeping mode &0.01 mW/s & \tabincell{c}{time of UE staying\\ in sleeping mode} \\ \hline
    $D_{\textbf{rx}}$& UE wakes up to listen to paging & 4 times/day &  \\ \hline
    $C$&  battery capacity &6500 J & \\ \hline
\end{tabular}
\end{center}
\end{table*}
In this work, system performance of the MTC services is investigated in dense urban environment as shown in Fig.~\ref{edm}, where a Madrid grid model is applied \cite{last}. The proposed environment model aligns well with the reality to generate meaningful and precise results. In this model, an urban environment is depicted with 3D visualization where each grid composes of one park and 15 buildings with different dimensions. The dimension of one Madrid grid is 387 meters in west-east direction and 552 meters in north-south direction. In order to achieve a cell radius of 866 meters, multiple replicas of Madrid grid are placed in the system level simulator. Moreover, building heights in the Madrid grids are uniformly distributed between 8 and 15 floors with a height of 3.5 meters per floor.
\subsection{Deployment Scenario}
A single macro BS with a cell radius of 886 meters is deployed in the Madrid grids, in order to achieve an inter site distance (ISD) of 1732m as defined in 3GPP \cite{lab5}\cite{ss}. The position of macro antennas is also plotted in Fig.~\ref{edm}. For the macro station, it operates in three cell sectors with a carrier frequency of 900 MHz and directional antennas are positioned with 120 degree difference from each other in the horizontal plane. 
\subsection{User deployment and traffic model}
In the coverage of the BS, 20 thousand sensor devices are randomly distributed inside buildings and are assumed to be static. An isotropic antenna is installed on each device at 1.5 meter height with a maximal transmission power of 23 dBm. Moreover, a report packet of 2000 bits with a periodicity of 24 reports per day is exploited in this work as traffic model for sensor devices.
\subsection{Channel model}
A 3D channel model proposed by 3GPP \cite{lab5} is applied here, in which the penetration loss through building floors and walls is taken into account. To characterize the channel in between the two ends of one D2D communication, channel models proposed in \cite{d2d} are applied. In \cite{d2d}, channel characters are captured in three different scenarios for indoor UEs, i.e.,
\begin{itemize}
\item two D2D ends are on the same floor in the same building;
\item two D2D ends are on the different floors in the same building;
\item two D2D ends are in different buildings.
\end{itemize}
\subsection{User device power consumption model}
In order to evaluation the power consumption of MTC devices, power consumption related parameters \cite{lab5}\cite{tt} are listed in Tab.~\ref{DC}. Please notice that, a new UE state called connected-inactive state is proposed to serve for 5G \cite{ii61}, and thus the duration of control plane establishment is calculated based on this new technology.
\section{Numerical results}\label{nr}
In this work, LTE technology is used in our work for modeling radio links and the mapping between SNR value and link capacity is performed by exploiting the results provided in \cite{final}. Thus, if a radio link experiences a SNR value lower than -7 dB, no data transmission is possible on this link. To reflect this aspect, availability is considered as the metric to be evaluated and its mathematical definition is given as:
\begin{equation}
\text{availability}=\frac{\text{number of UEs can be served}}{\text{total number of UEs}}.
\end{equation}
\begin{figure*}
\centering
\includegraphics[width=1.02\textwidth]{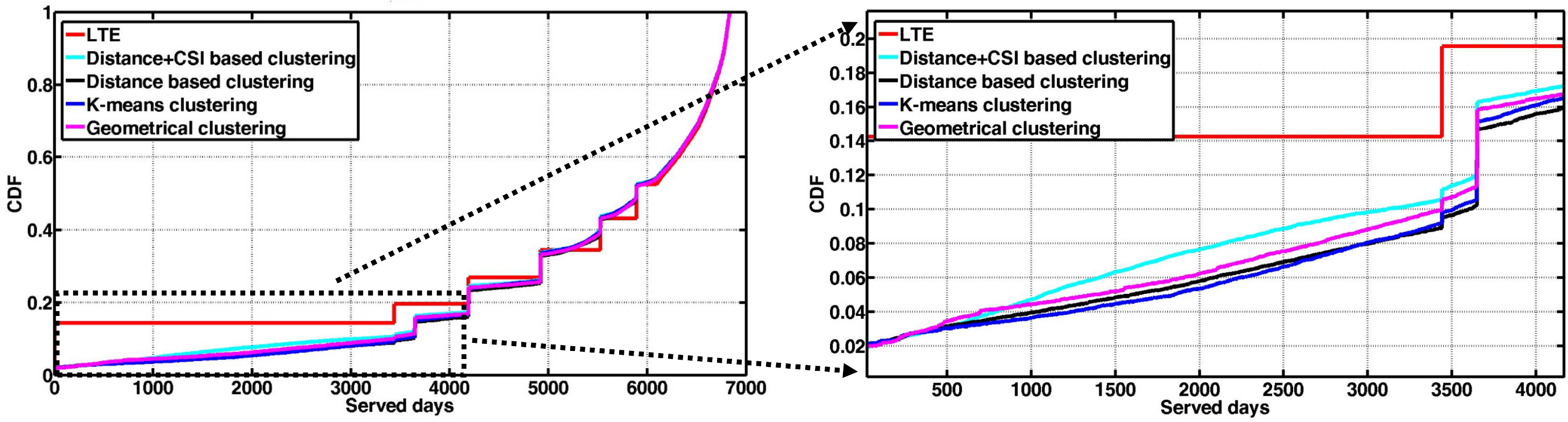}
\caption{system performance w.r.t device power consumption, $A_{\text{sector}}=40000\text{m}^2$}
\label{result1}
\end{figure*}
\begin{figure*}
\centering
\includegraphics[width=1.02\textwidth]{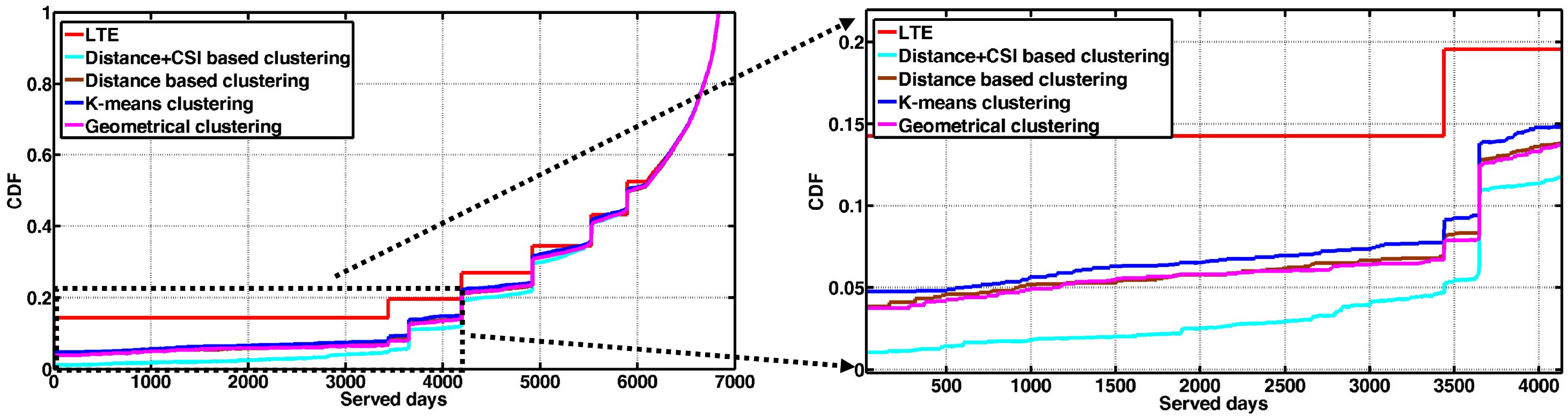}
\caption{system performance w.r.t device power consumption, $A_{\text{sector}}=2500\text{m}^2$}
\label{result2}
\end{figure*}
In other words, availability shows the ratio of users that can upload their packets to BS. Additionally, battery life of each UE is also inspected to calculate the ratio of users who can meet the battery life requirement of MTC services. Here, the target battery life is set to be ten years.\\
In the left hand of Fig.~\ref{result1}, the cumulative distribution function (CDF) of battery life of UEs is plotted, w.r.t. different clustering algorithms, as given in Sect.~\ref{tsm}. As a baseline scheme, the performance of LTE system is also drawn. As an input for geometrical clustering method, the area of a cluster is 40000 square meters. As an output from geometrical clustering method, number of clusters is further fed to other three clustering algorithms, in order to achieve a fair comparison among different algorithms. Moreover, a D2D pathloss value lower than 136dB is set to be the criterion for D2D connection setup, as stated in Sect.~\ref{signaling1}. Part of the CDF plot is zoomed in and shown at the right hand side of this figure, since it is the most interesting part for our inspection. As it can be seen from the figure, $14\%$ of MTC UEs can not transmit uplink reports to BS by using LTE technology, while only $2\%$ of UEs can not connect to BS if D2D communication is exploited. Thus, the availability is improved from $86\%$ to $98\%$. Moreover, $80.5\%$ of UEs can meet the battery life requirement of ten years (3650 days) in LTE system while this value can be improved to $90\%$ by exploiting D2D communication.\\
In Fig.~\ref{result2}, the same settings as in Fig.~\ref{result1} are applied, except that 2500 square meters is considered as the coverage area of one cluster for geometrical clustering method. Due to this fact, a larger number of clusters exist in the same coverage area of the BS and thus each cluster comprises a smaller number of UEs. In this sense, there is a lower probability of having feasible relay UE in each cluster. This is the reason why the availability values of three clustering algorithms are decreased to approximately $96\%$, compared with the values shown in Fig.~\ref{result1}. As an exception, the availability of the 'Distance+CSI based clustering' algorithm is improved to $99\%$. This is due to the fact that UEs with good channel conditions are selected as centroids in the initial step of the clustering algorithm. Moreover, by having more clusters, the coverage area of each cluster is reduced and the pathloss value between two D2D ends is smaller. Thus, the battery lives of UEs can be improved for all the applied D2D clustering schemes. For the 'Distance+CSI based clustering' algorithm, up to $95\%$ of UEs can meet the battery life requirement of ten years. 
\section{Conclusion}\label{conc}
As shown in this paper, a context-aware D2D communication can be applied to enhance the availability and improve device power consumption performance for MTC applications. Signaling schemes for D2D cluster formation and D2D communication are also provided to support the proposed context-aware D2D formation and TMS scheme. At the same time, the designed signaling schemes have also the advantages of less extra signaling overload compared with other schemes proposed in the literature. Moreover, the proposed concept is evaluated by a system level simulator and a large performance gain can be obtained by exploiting the proposed D2D communication.\\
\section*{Acknowledgment}
Part of this work has been performed in the framework of H2020 project METIS-II, which is funded by the European Union. The views expressed are those of the authors and do not necessarily represent the project. The consortium is not liable for any use that may be made of any of the information contained therein.

\begin{thebibliography}{1}
\bibitem{lab1}
Lien, S.Y.; Chen, K.C.; Lin,Y. \emph{Toward ubiquitous massive accesses in 3GPP machine-to-machine communications}, Communication Magazine. IEEE 2011, 49, 66-74.
\bibitem{lab2}
3GPP Technical Report 36.888. \emph{Study on provision of low-cost Machine-Type Communications (MTC) User Equipments (UEs) based on LTE (Release 12)}, June, 2013.
\bibitem{lab4}
Liang, J.M.; Chen J.J.; Cheng, H.H.; Tseng, Y.C. \emph{An Energy-Efficient Sleep Scheduling With QoS Considerationin 3GPP LTE-Advanced Networks for Internet of Things}, Emerging and Selected Topics in Circuits and Systems, IEEE Journal on 2013, 3, 13-22.
\bibitem{lab5}
3GPP Technical Report 45.820. \emph{Cellular system support for ultra-low complexity and low throughput Internet of Things (CIoT) (Release 13)}, November, 2015.
\bibitem{lab6}
Farhan Ahmad, Safdar Nawaz Khan Marwat, Yasir Zaki, Yasir Mehmood, Carmelita Goerg. \emph{Machine-to-machine Sensor Data Multiplexing using LTE-Advanced Relay Node for Logistics}, \url{https://files.nyu.edu/yz48/public/Final_LDIC_Farhan.pdf}
\bibitem{after6_1}
Ji Lianghai, A. Klein, N. Kuruvatti, H. D. Schotten, \emph{System Capacity Optimization Algorithm for D2D Underlay Operation}, in Proceedings of Workshop on 5G Technologies at IEEE International Conference on Communications (ICC), Sydney, Australia, June 2014.
\bibitem{after6_2}
Ji Lianghai, A. Klein, N. Kuruvatti, R. Sattiraju, H. D. Schotten, \emph{Dynamic Context-aware Optimization of D2D Communications}, in Proceedings of 2nd International Workshop on 5G Mobile and Wireless Communication System for 2020 and Beyond at IEEE 79th Vehicular Technology Conference (VTC-Spring), Seoul, Republic of Korea, May 2014
\bibitem{lab7}
M. Ji, G. Caire and A. F. Molisch. \emph{Wireless Device-to-Device Caching Networks: Basic Principles and System Performance}, in IEEE Journal on Selected Areas in Communications, vol. 34, no. 1, pp. 176-189, Jan. 2016.
\bibitem{lab8}
N.K. Pratas and P. Popovski. \emph{Underlay of low-rate machine-type D2D links on downlink cellular links}, 2014 IEEE International Conference on Communications Workshops (ICC), Sydney, NSW, 2014, pp. 423-428.
\bibitem{lab9}
N. K. Pratas and P. Popovski. \emph{Low-Rate Machine-Type Communication via Wireless Device-to-Device (D2D) Links}, arXiv preprint arXiv:1305.6783.
\bibitem{lab10}
3GPP meeting, RP-151948. \emph{New WI Proposal: D2D based MTC}, December 2015.
\bibitem{lab11}
Orsino, A., Araniti, G., Militano, L., Alonso-Zarate, J., Molinaro, A., Iera, A. (2016). \emph{Energy Efficient IoT Data Collection in Smart Cities Exploiting D2D Communications}, Sensors (Basel, Switzerland), 16(6), 836. \url{http://doi.org/10.3390/s16060836}
\bibitem{ITU}
ITU-R M.2135. \emph{Guidelines for evaluation of radio interface technologies for IMT-Advanced}, 2008.
\bibitem{last}
METIS, Deliverable 6.1, \emph{Simulation Guidelines}, October, 2013.
\bibitem{ss}
3GPP Technical Report 38.913. \emph{Study on Scenarios and Requirements for Next Generation Access Technologies}, October, 2016.
\bibitem{d2d}
3GPP, R1-132030. \emph{Channel models for D2D performance evaluation }, May, 2013.
\bibitem{tt}
Tuomas Tirronen, Anna Larmo, Joachim Sachs, Bengt Lindoff and Niclas Wiberg. \emph{Machine-to-machine communication with long term evolution with reduced device energy consumption}, Trans. Emerging Tel. Tech., 2013.
\bibitem{ii61}
METIS-II, Deliverable 6.1, \emph{Draft asynchronous control functions and overall control plane design}, June, 2016.
\bibitem{final}
Jar, M. and Fettweis, G. \emph{Throughput Maximization for LTE Uplink via Resource Allocation},   International Symposium on Wireless Communication Systems (ISWCS), 2012.

\end{thebibliography}
\end{document}